\documentclass[conference]{IEEEtran}
\IEEEoverridecommandlockouts
\usepackage{cite}
\usepackage{amsmath,amssymb,amsfonts}
\usepackage{algorithmic}
\usepackage{graphicx}
\usepackage{textcomp}
\usepackage{xcolor}
\usepackage{natbib} 
\usepackage{float}                
\usepackage{subfig}             
\usepackage{overpic}  
\usepackage{multicol}
\usepackage{multirow}
\usepackage{booktabs}
\usepackage{longtable}
\usepackage{bbding}
\usepackage{arydshln}
\usepackage{booktabs}
\usepackage{utfsym}
\usepackage{makecell}
\usepackage{verbatim}
\usepackage{CJKutf8}
\usepackage{natbib}
\setcitestyle{numbers,square}

\def\BibTeX{{\rm B\kern-.05em{\sc i\kern-.025em b}\kern-.08em
    T\kern-.1667em\lower.7ex\hbox{E}\kern-.125emX}}

\begin{document}
\begin{CJK}{UTF8}{gbsn}

\title{fMRI Exploration of Visual Quality Assessment\\
\thanks{
$^{\dag}$Equal contribution.\\
$^{*}$Corresponding author: Guangtao Zhai is with the Institute of Image Communication and Network Engineering, Shanghai Jiao Tong University, China, and also with the MoE Key Lab of Artificial Intelligence, AI Institute, Shanghai Jiao Tong University, China. E-mail: zhaiguangtao@sjtu.edu.cn. 
\\Yan Zhou, Department of Radiology, Renji Hospital, School of Medicine, Shanghai Jiao Tong University, 160 Pujian Road, Pudong New Area, Shanghai 200127, P.R. China.
Email: clare1475@hotmail.com}
}

\author{Yiming Zhang$^{1, \dag}$, Ying Hu$^{2, \dag}$, Xiongkuo Min$^{1, *}$, Yan Zhou$^{2, *}$, Guangtao Zhai$^{1, *}$\\
$^{1}$Institute of Image Communication and Network Engineering, Shanghai Jiao Tong University, China\\
$^{2}$Department of Radiology, Renji Hospital, School of Medicine,\\
Shanghai Jiao Tong University, Shanghai, China\\
\{ming\_zhang\_sjtu, minxiongkuo, zhaiguangtao\}@sjtu.edu.cn\\
clare1475@hotmail.com, hu19700204@163.com}
\vspace{2em}

\maketitle

\begin{abstract}
Despite significant strides in visual quality assessment, the neural mechanisms underlying visual quality perception remain insufficiently explored. This study employed fMRI to examine brain activity during image quality assessment and identify differences in human processing of images with varying quality.
Fourteen healthy participants underwent tasks assessing both image quality and content classification while undergoing functional MRI scans. The collected behavioral data was statistically analyzed, and univariate and functional connectivity analyses were conducted on the imaging data.
The findings revealed that quality assessment is a more complex task than content classification, involving enhanced activation in high-level cognitive brain regions for fine-grained visual analysis. Moreover, the research showed the brain's adaptability to different visual inputs, adopting different strategies depending on the input's quality. In response to high-quality images, the brain primarily uses specialized visual areas for precise analysis, whereas with low-quality images, it recruits additional resources including higher-order visual cortices and related cognitive and attentional networks to decode and recognize complex, ambiguous signals effectively.
This study pioneers the intersection of neuroscience and image quality research, providing empirical evidence through fMRI linking image quality to neural processing. It contributes novel insights into the human visual system's response to diverse image qualities, thereby paving the way for advancements in objective image quality assessment algorithms.
\end{abstract}


\section{Introduction}
In recent years, the field of visual neuroscience has witnessed remarkable advancements\cite{van2001mapping, grill2004human}. Functional magnetic resonance imaging (fMRI), as a non-invasive investigative technique, has emerged to strike an optimal balance between temporal and spatial resolution compared to electroencephalography (EEG) methods that are plagued by lower signal-to-noise ratios and inferior spatial precision\cite{logothetis2008we}. Consequently, fMRI has proven to be a productive tool in probing the intricacies of visual neurosciences.

Previous fMRI studies in vision have extensively relied on presenting subjects with specific types of visual stimuli such as faces, landscapes, objects, and abstract patterns to investigate the neural regions involved in processing visual information and their functional connectivity\cite{kanwisher1997fusiform, chang2017code, spiridon2006location, denys2004processing}. Notwithstanding these extensive explorations, most of these investigations have predominantly focused on the semantic content of the images rather than the influence of image quality on human visual comprehension. This area remains relatively unexplored within the context of fMRI-based visual research.\par

On the other hand, image quality assessment (IQA), as a pivotal subject within computer vision, centers on human perception of visual signals to reconcile resource allocation, such as channel bandwidth, system complexity, or production costs\cite{Zhai2020perceptual, lin2011perceptual}. This discipline bifurcates into subjective and objective assessment methods \cite{5404314}. Conventional subjective image quality assessment is grounded in psychophysical experimentation, where participants are required to provide judgments on the perceived image quality. The International Telecommunication Union (ITU) has established clear standards and definitions for these psychophysically-based perceptual quality tests \cite{Bt2002methodology, sector2008subjective}. These guidelines encompass not only the specifics of experimental parameters, which include viewing distance, ambient lighting conditions, stimulus presentation methodologies, categorical scaling techniques, and annotation of rating scales but also detailed prescriptions for how collected data should be processed and reported. The ITU recommendations ensure standardization across experiments, thereby enhancing the reliability and comparability of results from different studies in this field.\par

By contrast, certain researchers contend that psychophysiological tests relying on rating scales are inherently bound to conscious responses and may not efficaciously unravel the underlying perceptual and cognitive processes. Thus, alternative research avenues have emerged where investigators extract feature components from physiological signals, such as electroencephalography (EEG), to serve as objective indices of human perceptual quality, thereby circumventing the potential influence of subjective biases arising from high levels of cognitive engagement\cite{Engelke2016psychophysiology, scholler2012toward, bosse2017assessing, liu2019eeg}.\par

However, both subjective assessments based on psychometric scales and physiological signals are encumbered by the drawbacks of being time-consuming and costly, rendering them impractical for application to the vast quantities of images prevalent today. Thus, objective quality assessment methods hold greater research significance. Objective IQA endeavors to quantify image quality by grounding itself in the Mean Opinion Score (MOS) obtained from subjective assessments, emulating the human visual system's perceptional processes. In recent years, there has been remarkable progress in objective IQA algorithms, largely due to the powerful learning and adaptability capabilities of deep learning algorithms such as convolutional neural networks (CNNs) and transformers\cite{DBCNN, Su_2020_CVPR, ke2021musiq, yang2022maniqa, wang2022mstriq}. Nevertheless, these objective scoring algorithms, due to their inherent ``black box" nature of deep learning, lack interpretability, which impedes understanding the similarities and differences between their operations and the genuine perceptual processes within the human visual system. \par

To address the lacuna in visual neuroscience concerning the subject of image quality and to probe the neural underpinnings of image quality assessment, we have constructed an fMRI dataset dedicated to image quality assessment, conducted univariaty analyses and functional connectivity analyses. To our knowledge, this constitutes a pioneering endeavor at the intersection of neuroscience and quality assessment, where we aspire that this work will instigate novel insights into the human visual system's functioning and inform the design of future IQA algorithms.
\begin{figure*}[tp]	
  \centering
  \includegraphics[width=0.8\textwidth]{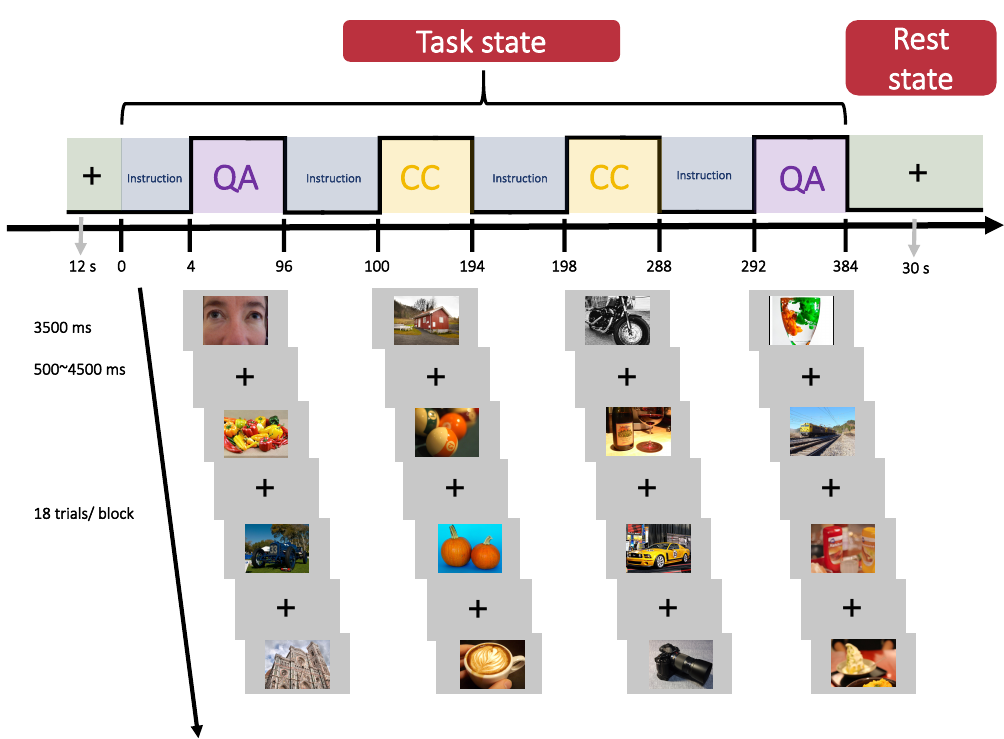} 
  \caption{Illustration of two main procedures of the proposed pruning method: train the decorators and get the mask for pruning in the split state, and retrain the model with the mask to recover its performance with merged convolutional layers.} 
  \label{fMRI_experiment} 
\end{figure*}
\section{MRI Experiment Methods}

\subsection{fMRI participates}
Eighteen, right-handed healthy university students participated in this study. 
Four participants were excluded for excessive motion, leaving 14 for analysis(7 males, ages 20-27, mean age=23.26).
Excessive motion was defined a priori as \textgreater 2 mm translation or \textgreater 2° rotation.
Participants had normal or corrected-to-normal visual acuity and no known neurological condition. The institutional review board at the Shanghai Jiao Tong University approved the experimental protocol, and all subjects gave written informed consent. After being informed about potential risks and screened by an institution's physician, subjects gave informed consent before participating. All data is processed anonymously. 

\subsection{Material}
All stimuli were classified into 9 categories according to content category (faces, objects, scene) and image quality (bad, neutral, excellent). All images are selected from Koniq-10k, an image quality assessment(IQA) database including images with authentic distortion and Mean Opinion Score (MOS) for their visual quality. All images were cropped to fit a $1024 \times 724$ pixels frame. These images were prepared for 2 tasks: image quality assessment and content classification. Each category for each task contains 18 different images. Each image was presented twice.

\subsection{fMRI experiment procedure}
The experimental process is divided into 8 runs, each containing 4 blocks. Participants in each block perform one of two tasks: quality assessment (bad, neutral, excellent) or content classification (faces, objects, scene), and each block contains 18 trials, as shown in Fig.\ref{fMRI_experiment}. Block order was fixed and counterbalanced within participants. On each trial, the stimulus was presented in the center of the screen for 3500ms including response time in random order, with a variable jitter time of 500–4500ms as inter-stimulus intervals. Nine categories of stimuli randomly appear, with the order and ISI designed using FreeSurfer optseq2. A fixation cross was presented for 12s at the beginning and for 30s at the end of each run. Scan data from the last 30 seconds of each run were treated as the resting state for subsequent analyses. All participants completed 8 runs.

\subsection{fMRI Data acquisition}
Visual stimulation is presented using the SINORAD SA-9939 Brain Functional Audiovisual Stimulation System with a 40" LCD screen for visual stimulation. The stimulation system is synchronized with the MRI system to provide a time reference for the stimulation task and MRI imaging. The stimulus program was written using E-Prime 3. \par
All MRI data were obtained on a 3T Siemens Prisma scanner at the Department of Radiology, Renji Hospital, School of Medicine, Shanghai Jiao Tong University, Shanghai, China. A gradient echo-planar imaging (EPI) sequence was used with the following parameters: TR=2000 ms, TE=30 ms, flip angle=90\textdegree, matrix size=64×64, FOV=192mm, slice thickness=2 mm, no inter-slice gap, 70 axial slices covering the whole brain, and Phase-encoding (PE) direction is Anterior-Posterior. In addition, T1-weighted three-dimensional (3D) structural images were acquired by using a MPRAGE sequence (TR=1800 ms, TE=2.28 ms, flip angle=8\textdegree, voxel-size=$1 mm\times1 mm\times1 mm$), T2-weighted three-dimensional (3D) structural images were acquired by using a TSE sequence (TR= 9560 ms, TE=90 ms, flip angle=150\textdegree, voxel-size=$2 mm\times2 mm\times2 mm$)

\subsection{Behavioural data analysis}
Participants' judgments in both tasks were fed back using buttons, and feedback and response time data were collected. We focus on the response time, which responds to the task's difficulty. We analyzed this data as part of our task analysis to determine whether image quality affects human image perception.
\begin{figure*}[htbp]	
  \centering
    \subfloat[QA task with different quality]{\includegraphics[width=0.25 \linewidth]{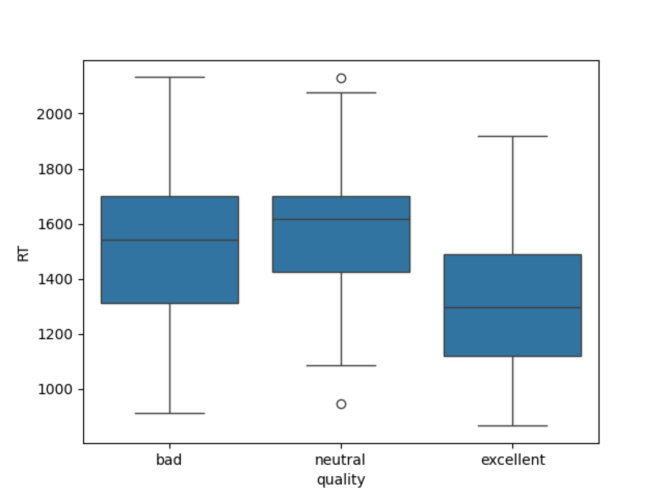}}
  \subfloat[CC task with different quality]{\includegraphics[width=0.25\linewidth]{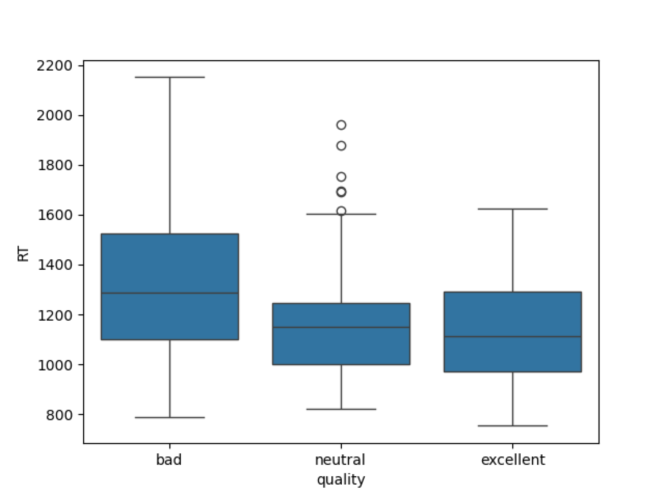}}
  \subfloat[QA task with different content]{\includegraphics[width=0.25 \linewidth]{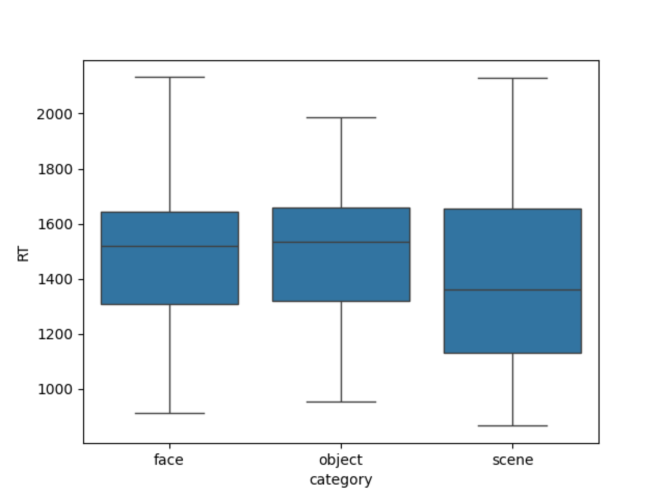}}
  \subfloat[CC task with different content]{\includegraphics[width=0.25 \linewidth]{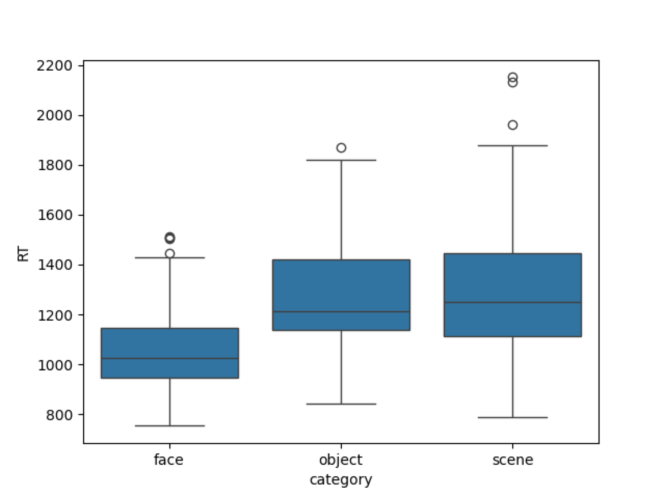}}
  \caption{Illustration of the distribution of average response times for images of different quality levels and content on the quality assessment and content classification task. } 
  \label{ANOVA} 
\end{figure*}

\subsection{fMRI data preprocessing}
All MRI data were converted into Brain Imaging Data Structure (BIDS) using dcm2niix (version v1.0.20220720). Results included in this manuscript come from preprocessing performed using \emph{fMRIPrep} 23.1.0 \cite{fmriprep1, fmriprep2} which is based on \emph{Nipype} 1.8.6 \cite{nipype1, nipype2}.\par

\textit{Preprocessing of B0 inhomogeneity mappings:}\quad
A \emph{B0} nonuniformity map (or \emph{fieldmap}) was estimated from the phase-drift map(s) measure with two consecutive GRE (gradient-recalled echo) acquisitions. The corresponding phase-map(s) were phase-unwrapped with \texttt{prelude} (FSL).\par
    
\textit{Anatomical data preprocessing :}\quad 
 Each T1-weighted (T1w) image was corrected for intensity non-uniformity (INU) with \texttt{N4BiasFieldCorrection} \cite{n4}, distributed with ANTs \cite{ants}, and used as T1w-reference throughout the workflow. The T1w-reference was then skull-stripped with a \emph{Nipype} implementation of the \texttt{antsBrainExtraction.sh} workflow (from ANTs), using OASIS30ANTs as the target template. Brain tissue segmentation of cerebrospinal fluid (CSF), white matter (WM), and gray-matter (GM) was performed on the brain-extracted T1w using \texttt{fast} \cite{fsl_fast}. Brain surfaces were reconstructed using \texttt{recon-all} \cite{fs_reconall}, and the brain mask estimated previously was refined with a custom variation of the method to reconcile ANTs-derived and FreeSurfer-derived segmentations of the cortical gray-matter of Mindboggle \cite{mindboggle}. Volume-based spatial normalization to one standard space (MNI152NLin2009cAsym) was performed through nonlinear registration with \texttt{antsRegistration} (ANTs (version unknown)), using brain-extracted versions of both T1w reference and the T1w template. The following templates were selected for spatial normalization and accessed with \emph{TemplateFlow} \cite{templateflow}: \emph{ICBM 152 Nonlinear Asymmetrical template version 2009c} \cite{mni152nlin2009casym}.\par
 
\textit{Functional data preprocessing:}\quad 
For each of the 8 BOLD runs found per subject (across all tasks and sessions), the following preprocessing was performed. First, a reference volume and its skull-stripped version were generated using a custom methodology of \emph{fMRIPrep}. Head-motion parameters with respect to the BOLD reference (transformation matrices, and six corresponding rotation and translation parameters) are estimated before any spatio-temporal filtering using \texttt{mcflirt} \cite{mcflirt}. The estimated \emph{fieldmap} was then aligned with rigid-registration to the target EPI (echo-planar imaging) reference run. The field coefficients were mapped onto the reference EPI using the transform. BOLD runs were slice-time corrected to 0.961s (0.5 of slice acquisition range 0s-1.92s) using \texttt{3dTshift} from AFNI \cite{afni}. The BOLD reference was then co-registered to the T1w reference using \texttt{bbregister} (FreeSurfer) which implements boundary-based registration \cite{bbr}. Co-registration was configured with six degrees of freedom. 
The three global signals are extracted within the CSF, the WM, and the whole-brain masks. 
 A high-pass filter with a cut-off of 128s was applied to remove low-frequency noise. The BOLD time-series were resampled into standard space, generating a \emph{preprocessed BOLD run in MNI152NLin2009cAsym space}. First, a reference volume and its skull-stripped version were generated using a custom methodology of \emph{fMRIPrep}. All resamplings can be performed with \emph{a single interpolation step} by composing all the pertinent transformations. 
Gridded (volumetric) resamplings were performed using \texttt{antsApplyTransforms} (ANTs), configured with Lanczos interpolation to minimize the smoothing effects of other kernels. \cite{lanczos}. Non-gridded (surface) resamplings were performed using \texttt{mri\_vol2surf} (FreeSurfer). \par

After preprocessed by fMRIPrep, the functional data for the 576 trials were analyzed and extracted using the following steps. 
For univariate analysis and functional connectivity analysis, the functional data were spatially smoothed with an isotropic 4 mm full-width-half-maximum (FWHM) Gaussian kernel. 
Then the data before and after being smoothed from each run were all entered into the general linear model (GLM) to regress nuisance variables out of the data (i.e. denoising operation). There were four kinds of nuisance variables including motion correction parameters, white matter(WM) and cerebral spinal fluid(CSF) BOLD time series, and scrubbing (i.e. identified outlier scans). All of the nuisance variables were confounds extracted in the fMRIPrep analysis stream. Furthermore,  a temporal band-pass filter of 0.008-0.09Hz was applied to reduce the low-frequency drift derived from head motions or physiological sources.

\subsection{fMRI data analysis}
\textit{Univariate analysis :}\quad 
The classical univariate statistical analysis was performed on SPM12. In the first level analysis, we used the quality assessment task and content classification task as regressors of interest to explore brain mechanisms for quality assessment tasks and used high-quality and low-quality as regressors of interest to find brain regions sensitive to image quality. A high-pass filter of 128s and an AR(1) model were used for signal drift correction and serial correlations respectively. The statistically significant thresholds for reported group-level analysis results were set at a voxel-level $p \textless 0.001$ and FDR corrected ($p \textless 0.05$).\par
\textit{Functional connectivity analysis :}\quad 
The functional connectivity analysis was performed using the CONN toolbox. Given the lack of research on image quality-sensitive regions in the human brain, and to avoid the limitations imposed by imperfect a priori knowledge, we used an ROI-to-ROI approach to explore the complete QA task-related brain network. The ROIs were defined by combining FSL Harvard-Oxford atlas cortical and subcortical areas with an extended set of classical networks (138 ROIs in total). 
To measure the level of task-modulated effective connectivity between two ROIs, we employed the Generalized Psycho-Physiological Interaction (gPPI) approach with bivariate regression for the first-level analysis. For the group-level analysis, the Spatial Pairwise Clustering (SPC) statistics were used for Cluster-level inferences. The thresholds of statistical significance were set at a cluster level FDR corrected $p \textless 0.05$ and voxel level uncorrected $p \textless 0.01$.

\begin{figure*}[htbp]	
  \centering
  \subfloat[Quality Assessment vs Content Classification]
  {\includegraphics[width=0.5 \linewidth]{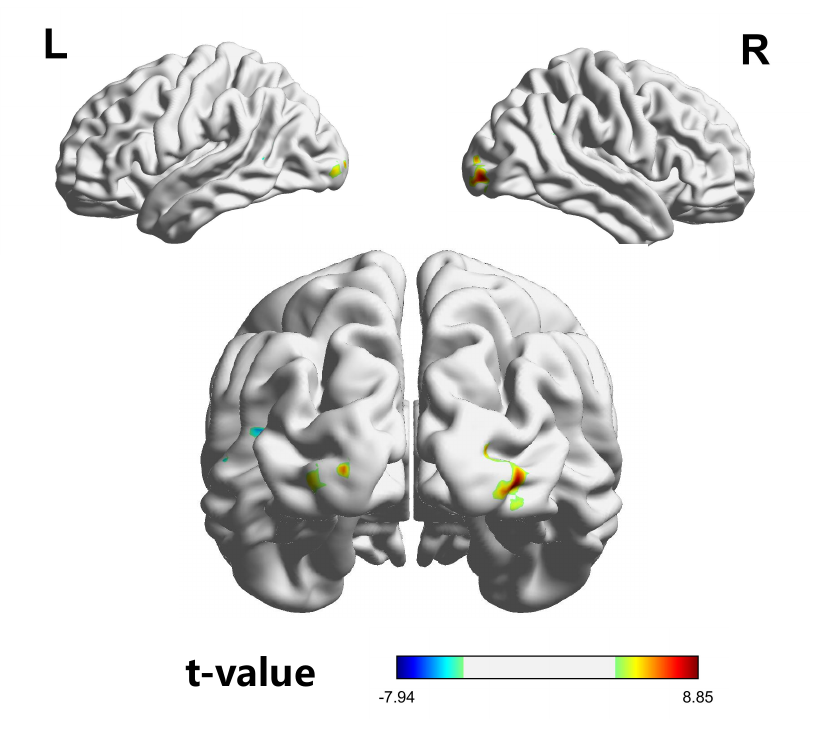}}
  \subfloat[High-quality vs Low-quality]
  {\includegraphics[width=0.5\linewidth]{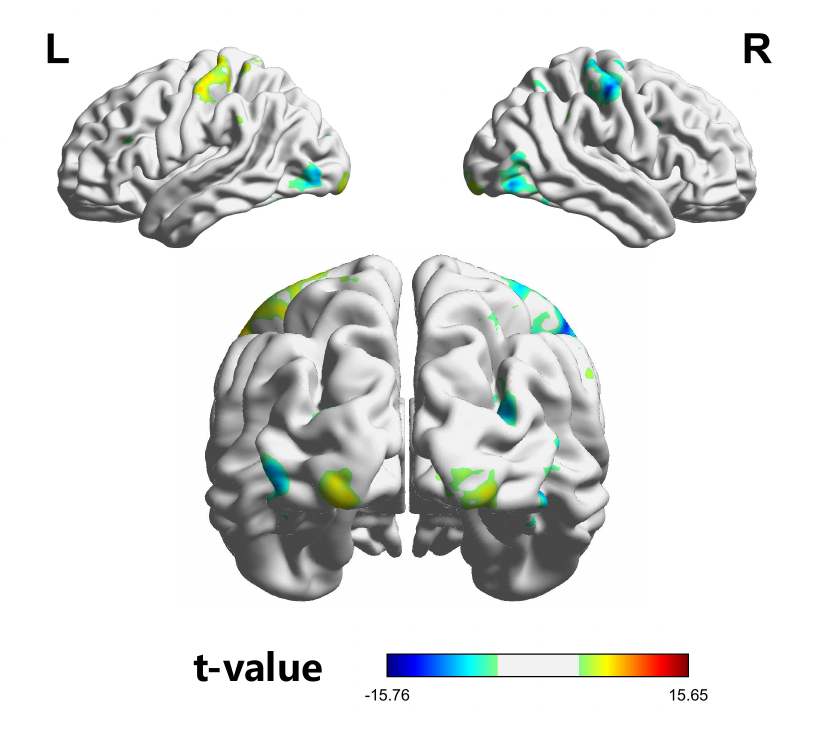}}
  \caption{The effects of task and quality condition on blood-oxygen-level-dependent (BOLD) activity. (\textbf{a}) brain regions showing increased and decreased activities during the QA task compared with the CC task. (\textbf{b}) Regions of the brain show increased and decreased activity when viewing high-quality images compared to low-quality images.} 
  \label{spm_brain} 
\end{figure*}

\section{MRI Results}
\subsection{Behavioural results}
First, we conducted an ANOVA with image quality(bad, neutral, excellent) as within-trial variables, and average response time (RT) as the dependent variable.
It revealed a significant main effect of image quality on the quality assessment and content classification task, $F(2, 286)=32.61, p \textless 0.0001; F(2, 286)=16.85, p \textless 0.0001$ respectively. The box plots illustrating the distribution of the average response time required for the quality assessment and content classification task for images of different quality levels are shown in Fig. \ref{ANOVA}. The box plots show that participants took longer to recognize the semantics of low-quality images compared to medium and high-quality images. \par
No significant interaction between content categories and response time on the quality assessment task was observed, $F(2, 286)=4.099, p=0.018$. However, a significant main effect of content category on the content classification task was found, $F(2, 286)=31.92,  p \textless 0.0001$. The box plots illustrating the distribution of the average response time required for the quality assessment and content classification task for images of different content categories are shown in Fig. \ref{ANOVA}. The box plots show that humans are significantly faster at judging faces in the content classification task compared to objects and scenes.\par
Furthermore, we have scrutinized the mean RTs for both the QA and CC tasks, which were found to be 1.468 seconds and 1.213 seconds, respectively, across all trials and participants. It is noteworthy that the task type exerted a significant influence on the mean response latency, as evidenced by an F-statistic of F(1,287) = 140.43, with $p \textless 0.0001$, indicating that the QA task is indeed more demanding compared to the CC task.

\begin{table*}
    \centering
\caption{Regions activated and inhibited under the quality assessment task in contrast to content recognition. Results of univariate analyses from SPM ($p \textless 0.001$, FDR corrected at voxel level ($p \textless 0.05$)) shown below include cluster sizes, peak level t-scores, and the MNI coordinates of each peak and their corresponding atlas labels based on Neuromorphometrics and Brodmann’s area (BA) from whole-brain analysis. For large clusters covering multiple brain regions, we also report the main structures contained. }
    \label{QA-OC_2}
    \begin{tabular}{c|ccccccc}
         \toprule
         Contrast & Cluster size & \makecell{Main structure\\in cluster}
         & \makecell{Peak MNI\\coordinates}  
         & Brain region & Hemisphere &BA & t-score 
         \\ 
         \midrule
         \multirow{12}{*}{QA\textgreater CC} 
         &\multirow{3}{*}{2440}  
         & \multirow{3}{*}{\makecell{Occipital Lobe, Lingual Gyrus, \\Cuneus, Fusiform Gyrus,\\ Inferior Temporal Gyrus}}
         & (10,-96,14) & Cuneus & R & 18 & 12.40 \\
         & & &(18,-94,10) & Middle Occipital Gyrus & R & -- & 11.85\\
         & & &(8,-92,-2) &  Lingual Gyrus & R &17 & 11.71\\
         
         \cdashline{2-8}
         &\multirow{3}{*}{2036}  
         & \multirow{3}{*}{\makecell{Occipital Lobe, Lingual Gyrus, \\Cuneus, Fusiform Gyrus}}
         & (-26,-64,-10) &  Lingual Gyrus & L & -- & 11.36 \\
         & & &(-24,-98,12) & Middle Occipital Gyrus & L & -- & 11.30\\
         & & &(-28,-78,-10) &  Lingual Gyrus & L &  & 11.06\\ 
        
        \cdashline{2-8}
         & 76 & -- & (58,12,20) & Inferior Frontal Gyrus & R & 45 & 6.52 \\
         & 19 & -- & (-12,28,62) &  Superior Frontal Gyrus & L & 6 & 5.69 \\    
         & 22 & -- & (38,-2,12) &  Insula & R & 13 & 5.50 \\
         & 11 & -- & (-58,0,-24) & Middle Temporal Gyrus & L & 21 & 5.46 \\
         & 21 & -- & (66,-10,22) & Postcentral Gyrus & R & 43 & 5.18 \\
         & 18 & -- & (-50,8,32) &  Inferior Frontal Gyrus & L & -- & 4.57 \\
         
         \midrule
         
         \multirow{11}{*}{CC\textgreater QA} 
         &\multirow{3}{*}{651}  
         & \multirow{3}{*}{\makecell{Cerebrum, Postcentral Gyrus, \\Precuneus, Cingulate Gyrus,\\ Superior Parietal Lobule}}
         & (14,-34,40) & Cingulate Gyrus & R & -- & 7.61 \\
         & & &(-2,-48,60) &  Precuneus & L & -- & 7.52\\
         & & &(8,-92,-2) & Cingulate Gyrus & R & -- & 7.07\\
         
         \cdashline{2-8}
         &\multirow{3}{*}{123}  
         & \multirow{3}{*}{\makecell{Cuneus, Precuneus, \\Occipital Lobe}}
         & (-6,-38,36) & Precuneus  & L & 7 & 6.99 \\
         & & &(-14,-84,34) &  Cuneus & L & 19 & 5.36\\
         & & &(8,-68,36) &  Precuneus & R & 7 & 5.07\\
        
        \cdashline{2-8}
         & 62 & -- & (-40,38,36) & Superior Frontal Gyrus & L & 9 & 7.74 \\
         & 27 & -- & (14,-58,40) & Precuneus & R & -- & 7.67 \\
         & 38 &  Angular & (52,-66,26) & Middle Temporal Gyrus & R & 39 & 6.27 \\
         & 21 & -- & (30,20,60) & Middle Frontal Gyrus & R & 6 & 5.88 \\
         & 62 & -- & (-36,-74,42) & Inferior Parietal Lobule & L & 19 & 5.85 \\
         
         \bottomrule
    \end{tabular}
\end{table*}

\begin{table*}
    \centering
\caption{Regions activated and inhibited under the high-quality condition in contrast to low-quality condition. Results of univariate analyses from SPM ($p \textless 0.001$, FDR corrected at voxel level ($p \textless 0.05$)) shown below include cluster sizes, peak level t-scores, and the MNI coordinates of each peak and their corresponding atlas labels based on Neuromorphometrics and Brodmann’s area (BA) from whole-brain analysis. For large clusters covering multiple brain regions, we also report the main structures contained.  }
    \label{tab_good_bad}
    \begin{tabular}{c|ccccccc}
         \toprule
         Contrast & Cluster size & \makecell{Main structure\\in cluster}
         & \makecell{Peak MNI\\coordinates}  
         & Brain region & Hemisphere &BA & t-score 
         \\ 
         \midrule
         \multirow{18}{*}{\makecell{High-quality\textgreater \\Low-quality}} 
         &\multirow{3}{*}{1946}  
         & \multirow{3}{*}{\makecell{Postcentral Gyrus, Precuneus, \\Superior Parietal Lobule}}
         & (-40,-24,44) & Postcentral Gyrus & \multirow{3}{*}{L} & 4 & 15.65 \\
         & & &(-36,-24,58) &   \multirow{2}{*}{Precentral Gyrus} & & -- & 11.48\\
         & & &(-38,-22,66) & & &3 & 10.28\\
         
         \cdashline{2-8}
         &\multirow{3}{*}{139}  
         & \multirow{3}{*}{\makecell{Postcentral Gyrus, Insula, \\Superior Temporal Gyrus}}
         & (-46,-22,22) &   Insula & \multirow{3}{*}{L} & -- & 9.51 \\
         & & &(-56,-26,22) & Postcentral Gyrus &  & -- & 5.90\\
         & & &(-50,-20,8) &  Superior Temporal Gyrus &  & -- & 5.22\\

        \cdashline{2-8}
         &\multirow{3}{*}{445}  
         & \multirow{3}{*}{\makecell{Cuneus, Middle Occipital Gyrus,\\ Lingual Gyrus, BA17, \\Inferior Occipital Gyrus,  BA18}}
         & (30,-98,-6) &  Middle Occipital Gyrus & \multirow{3}{*}{R}  & 18 & 9.25 \\
         & & &(14,-94,-10) & Lingual Gyrus & & -- & 7.41\\
         & & &(26,-92,-14) &   Inferior Occipital Gyrus & & -- & 5.22\\ 

        \cdashline{2-8}
         &\multirow{3}{*}{540}  
         & \multirow{3}{*}{\makecell{Cuneus, Middle Occipital Gyrus,\\ Lingual Gyrus, BA17, \\Inferior Occipital Gyrus,  BA18}}
         & (-24,-96,-10) &  Lingual Gyrus & \multirow{3}{*}{L}  & -- & 8.20 \\
         & & &(-12,-100,-10) & Lingual Gyrus & & 18 & 7.76\\
         & & &(-24,-100,-10) & Middle Occipital Gyrus & & -- & 7.68\\ 

        \cdashline{2-8}
         &\multirow{3}{*}{154}  
         & \multirow{3}{*}{\makecell{Paracentral Lobule, Parietal Lobe, \\Precuneus, Cingulate Gyrus}}
         & (-14,-38,52) &  Paracentral Lobule & \multirow{3}{*}{L}  & 5 & 6.81 \\
         & & &(-6,-30,42) &  Cingulate Gyrus &  & 31 & 5.69\\
         & & &(-4,-22,46) & Paracentral Lobule &  & 31 & 5.53\\ 
        
        \cdashline{2-8}
         & 158 & -- & (62,-42,36) &  Inferior Parietal Lobule & R & 40 & 7.18\\
         & 96 & -- & (-64,-38,34) &  Inferior Parietal Lobule & L & 40 & 6.71\\
         & 20 & -- & (-44,4,-6) &  Insula & L & -- & 5.99\\
         & 29 & -- & (14,-32,40) & Cingulate Gyrus & R & 31 & 5.95\\

         \midrule
         
         \multirow{23}{*}{\makecell{Low-quality\textgreater \\High-quality}} 
         &\multirow{4}{*}{2826}  
         & \multirow{4}{*}{\makecell{Postcentral Gyrus, Superior Parietal Lobule,\\ Middle Frontal Gyrus, Precentral Gyrus,\\ BA4, BA3, BA6, BA40, BA19}}
         & (34,-26,62) & \multirow{3}{*}{Precentral Gyrus} & \multirow{3}{*}{R}  & 4 & 15.76 \\
         & & &(50,-14,52) & & & -- & 14.85\\
         & & &(40,-20,62) & & & 4 & 14.05\\ 
         &&&&&&&\\

         \cdashline{2-8}
         &\multirow{4}{*}{679}  
         & \multirow{4}{*}{\makecell{Middle Occipital Gyrus, BA37, BA39,\\Middle Temporal Gyrus, Fusiform,\\Inferior Temporal Gyrus }}
         & (-48,-84,2) & \multirow{3}{*}{Middle Occipital Gyrus} & \multirow{3}{*}{L} & -- & 12.62 \\
         & & &(-46,-76,4) & & & 19 & 9.60\\
         & & &(-44,-76,-4) & & & -- & 8.29\\
         &&&&&&&\\

         \cdashline{2-8}
         &\multirow{3}{*}{918}  
         & \multirow{3}{*}{\makecell{Middle Occipital Gyrus, BA37, BA19,\\Inferior Temporal Gyrus, Fusiform Gyrus,\\Middle Temporal Gyrus}}
         & (42,-74,-4) & Inferior Temporal Gyrus & \multirow{3}{*}{R} & -- & 11.53 \\
         & & &(50,-72,0) & Middle Occipital Gyrus & & 37 & 9.11\\
         & & &(52,-72,12) &  Middle Temporal Gyrus & & -- & 8.23\\
       
         \cdashline{2-8}
         &\multirow{3}{*}{300}  
         & \multirow{3}{*}{\makecell{Inferior Frontal Gyrus, \\Middle Frontal Gyrus,\\BA8, BA9}}
         & (44,4,28) &  Inferior Frontal Gyrus  & \multirow{3}{*}{R} & -- & 8.18 \\
         & & &(54,8,24) & Inferior Frontal Gyrus & & -- & 7.71\\
         & & &(54,18,36) & Middle Frontal Gyrus & & -- & 5.99\\

         \cdashline{2-8}
         &\multirow{2}{*}{153}  
         & \multirow{2}{*}{\makecell{Superior Parietal Lobule, \\Inferior Parietal Lobule}}
         & (-34,-64,54) & Superior Parietal Lobule & \multirow{2}{*}{L}  & 7 & 7.72 \\
         & & &(-36,-54,52) &  Inferior Parietal Lobule & & -- & 5.22\\

         \cdashline{2-8}
         &\multirow{2}{*}{203}  
         & \multirow{2}{*}{\makecell{Middle Frontal Gyrus ,\\Inferior Frontal Gyrus}}
         & (-52,6,28) & Inferior Frontal Gyrus &\multirow{2}{*}{L} & -- & 7.36 \\
         & & &(-50,6,44) &  Middle Frontal Gyrus & & -- & 5.30\\

         \cdashline{2-8}
         &\multirow{3}{*}{164}  
         & \multirow{3}{*}{\makecell{Middle Occipital Gyrus,\\Superior Occipital Gyrus\\BA19}}
         & (-24,-88,26) & \multirow{3}{*}{Cuneus} & \multirow{3}{*}{L} & -- & 6.06 \\
         & & &(-26,-76,20) & & & -- & 5.93\\
         & & &(-20,-90,16) & & & -- & 4.82\\

        \cdashline{2-8}
         & 164 & -- & (-24,-88,26) & Cuneus  & R & 7 & 6.06 \\
         & 34 & -- & (40,-2,12) &  Insula & R & -- & 6.41 \\
         & 67 & -- & (52,30,22) & Middle Frontal Gyrus & R & 46 & 6.13 \\
         & 56 & -- & (36,-12,16) &  Insula & R & 13 & 6.05 \\
         
         \bottomrule
    \end{tabular}
\end{table*}

\begin{figure*}[htbp]	
  \centering
  \subfloat[Positive]
  {\includegraphics[width=0.5 \linewidth]{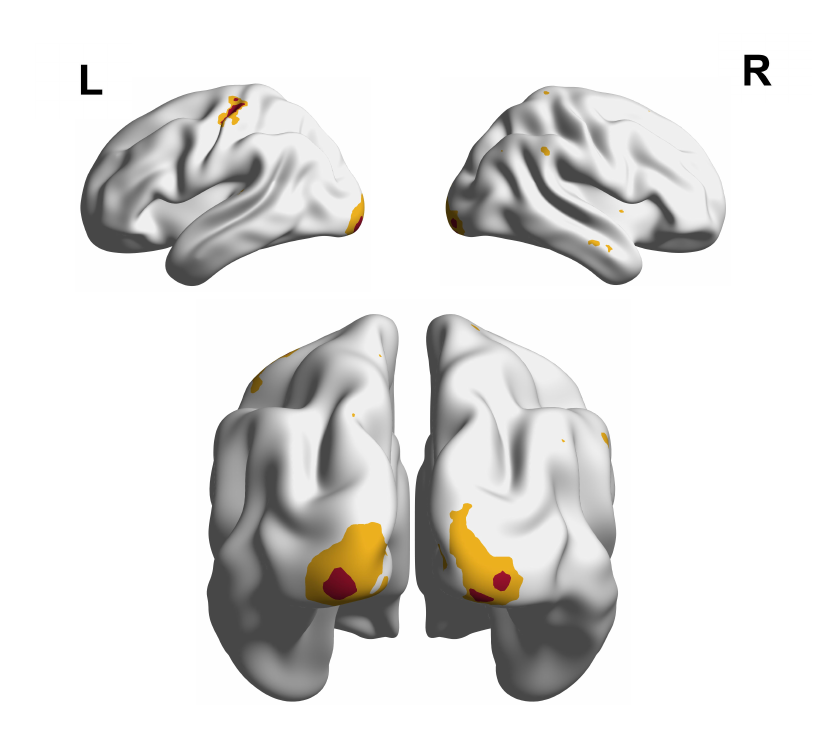}}
  \subfloat[Negative]
  {\includegraphics[width=0.5\linewidth]{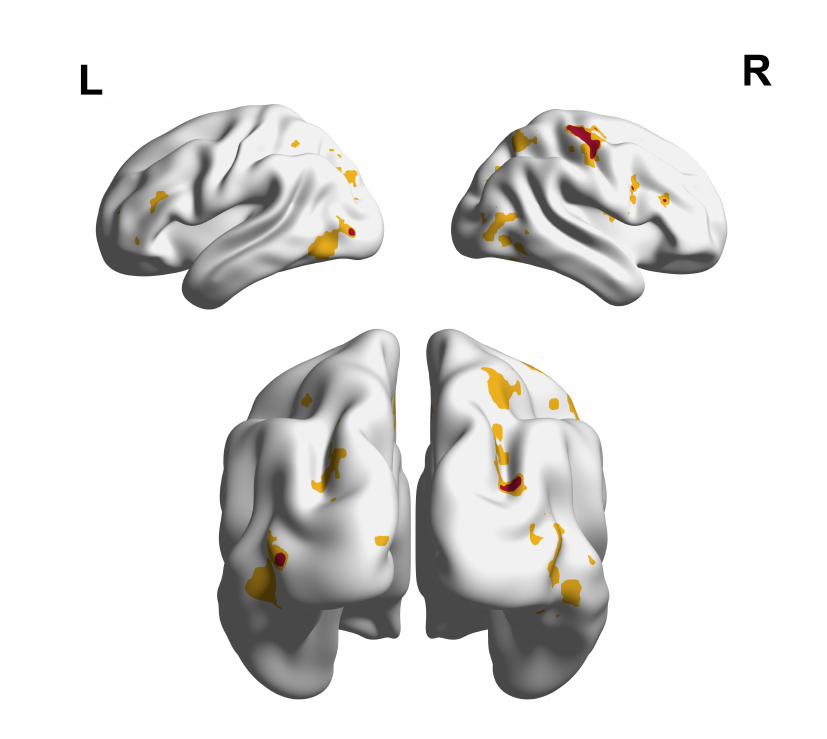}}\\
    \subfloat[Lingual Gyrus]
  {\includegraphics[width=0.25 \linewidth]{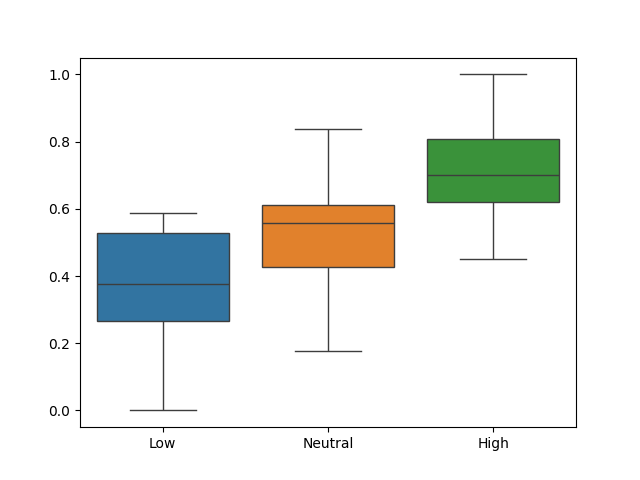}}
    \subfloat[Middle Occipital Gyrus]
  {\includegraphics[width=0.25 \linewidth]{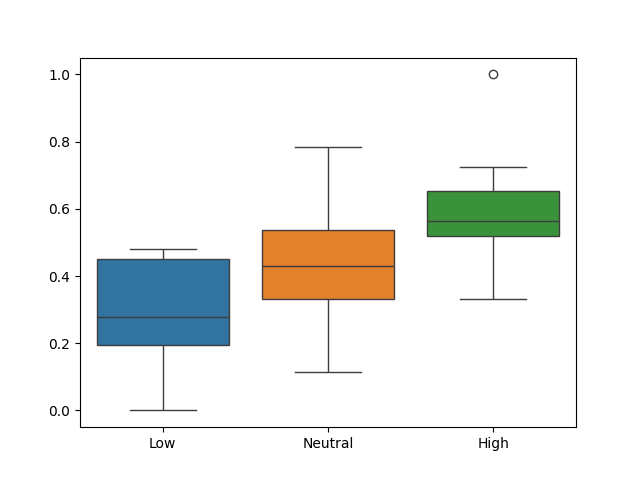}}
  \subfloat[Superior Occipital Gyrus]
  {\includegraphics[width=0.25 \linewidth]{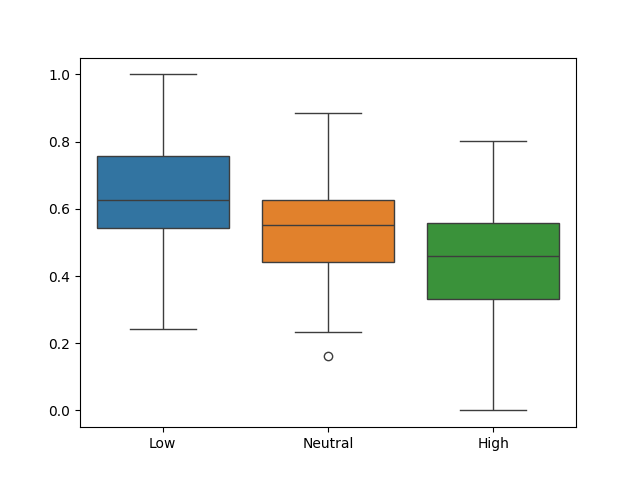}}
   \subfloat[Middle Frontal Gyrus]
  {\includegraphics[width=0.25 \linewidth]{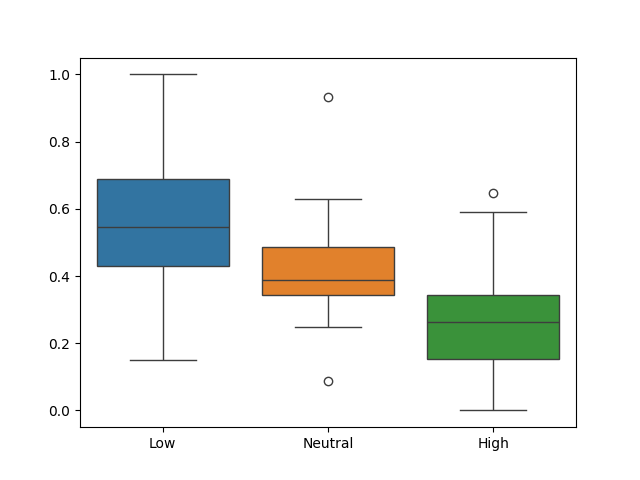}}
  \caption{Brain regions demonstrating significant activation across both ``High-quality vs Neutral-quality" and ``Neutral-quality vs Low-quality" contrasts scenarios. (a) and (b) respectively illustrate regions of positive and negative activations under the two distinct contrasts. The hue of yellow has been employed to highlight brain territories that exhibit activation solely within one contrast, whereas the color magenta represents regions showing consistent activation across both contrasts. The criterion for statistical significance, denoting activation, has been conservatively established at a threshold level of $p \textless 0.001$, FDR corrected at voxel level ($p \textless 0.05$). (c) - (f) respectively illustrate the distribution of Blood Oxygen Level Dependent (BOLD) signal intensity within subregions of the Lingual Gyrus, Middle Occipital Gyrus, Superior Occipital Gyrus, and Middle Frontal Gyrus across different image quality levels for each of the 14 participants. The vertical axis in these plots has been normalized to allow comparison, with the specific demarcation of these subregions derived from the common activation regions identified in (a) and (b).} 
  \label{gradual} 
\end{figure*}

\subsection{fMRI univariate analyses}
\subsubsection{Brain regions for quality assessment task}
To unravel the brain mechanisms underlying image quality assessment, we first performed univariate analyses to explore the brain regions activated in the quality assessment task. 
We focus on which brain regions are more sensitive in the low-level quality assessment task in contrast to the usual high-level task focusing on image semantics, so the contrast ``Quality Assessment(QA) vs Content Classification(CC)" was conducted, as shown in Fig.\ref{spm_brain}. The statistical results are presented in detail in Table \ref{QA-OC_2}. In particular, we observed significantly increased activity in the brain regions in the visual pathway, such as the middle occipital gyrus, lingual, fusiform, cuneus, and inferior temporal gyrus in the QA task compared to the CC task. In addition, increased activity was observed in regions such as the bilateral inferior frontal gyrus, the left superior frontal gyrus, the right insula, and the right postcentral gyrus. \par
Moreover, we found reduced activity in the QA task compared to the CC task in the right cingulate gyrus and bilateral precuneus, left superior frontal gyrus, and right middle temporal gyrus\par

\subsubsection{Brain regions for different quality images}
In addition to exploring the brain regions activated by the quality assessment task, we were also curious to investigate whether images of different quality would activate different regions of the human brain, thus exploring the physiological basis of quality of experience (QoE). The contrast between high-quality and low-quality conditions was conducted. The statistical results are presented in detail in Table \ref{tab_good_bad}.
First, we can observe conjugate activation of high and low quality in the precentral gyrus, i.e. high and low-quality leads to positive activation in the left and right precentral gyrus, respectively. This is because we specify that the button representing low quality is on the left and the button representing high quality is on the right. So the activation here is not related to the image quality itself. Conjunctive activation of high and low quality also can be observed in the insula, i.e. high and low-quality images led to positive activation in the left and right insula, respectively.
In contrast ``High-quality \textgreater Low-quality", some subregions in the visual regions, such as the lingual gyrus, inferior occipital gyrus, and middle occipital gyrus, responded at a higher level to high-quality images, mainly involving regions BA17 and BA18. In addition, significant activation can also be observed in the bilateral inferior parietal lobule and bilateral Cingulate Gyrus, as shown in Fig.\ref{spm_brain}. 
In contrast ``Low-quality \textgreater High-quality", significant activation can be observed in the subregion of bilateral fusiform, middle occipital gyrus, and superior occipital gyrus, mainly involving regions BA19 and BA37. Other activations are found mainly in the bilateral middle temporal gyrus, inferior temporal gyrus, inferior frontal gyrus, middle frontal gyrus, and the left superior parietal lobule and inferior parietal lobule.

\subsubsection{Further analyses of quality-related regions}
Given the established responsiveness of specific cerebral regions to variations in stimulus quality, our study delved into elucidating the brain areas exhibiting either a positive or negative correlation with BOLD signal intensity as a function of quality. In pursuit of this objective, we performed distinct univariate analyses comparing "High-quality vs Neutral-quality" conditions and separately, "Neutral-quality vs Low-quality" conditions. This was succeeded by pinpointing the common brain regions demonstrating consistent significant positive or negative activation across both contrast scenarios. Thereafter, we scrutinized the GLM-estimated beta parameters within these overlapping regions across the spectrum of quality levels, encompassing High-quality, Neutral-quality, and Low-quality conditions, as shown in Fig.\ref{gradual}.

An examination of the results indicates that there exists a pattern of elevated BOLD responses in sub-regions associated with the lingual gyrus and middle occipital gyrus(mainly BA18) as image quality ascends. On the other hand, a subset of regions within the superior occipital gyrus and middle frontal gyrus display an inverted phenomenon, wherein a decrement in visual input quality corresponds to an enhancement of BOLD signal responsiveness.

\begin{figure*}[htbp]	
  \centering
  \subfloat[Quality Assessment Task]
  {\includegraphics[width=1 \linewidth]{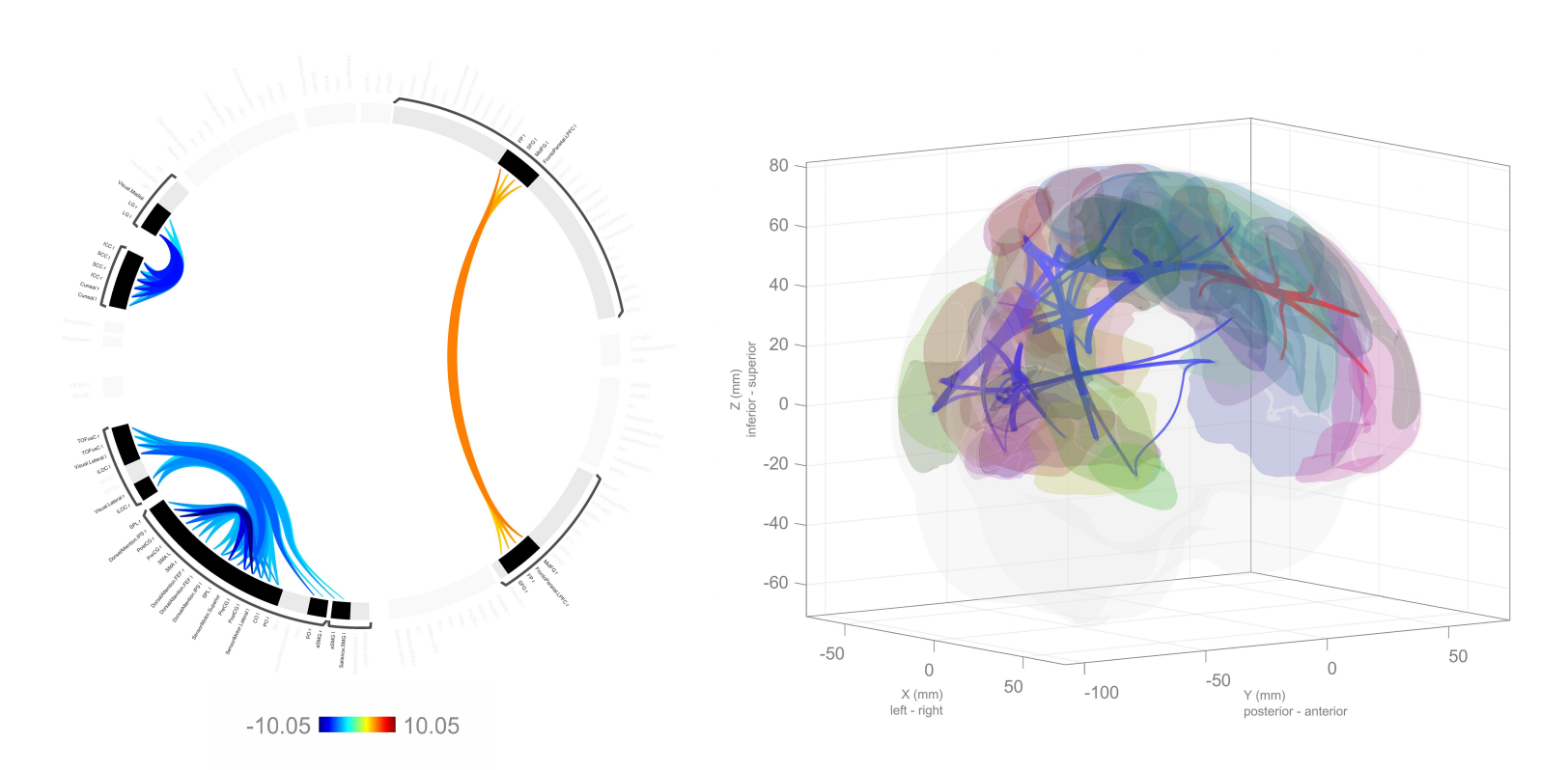}}\\
  \subfloat[Content Classification Task]{\includegraphics[width=1\linewidth]{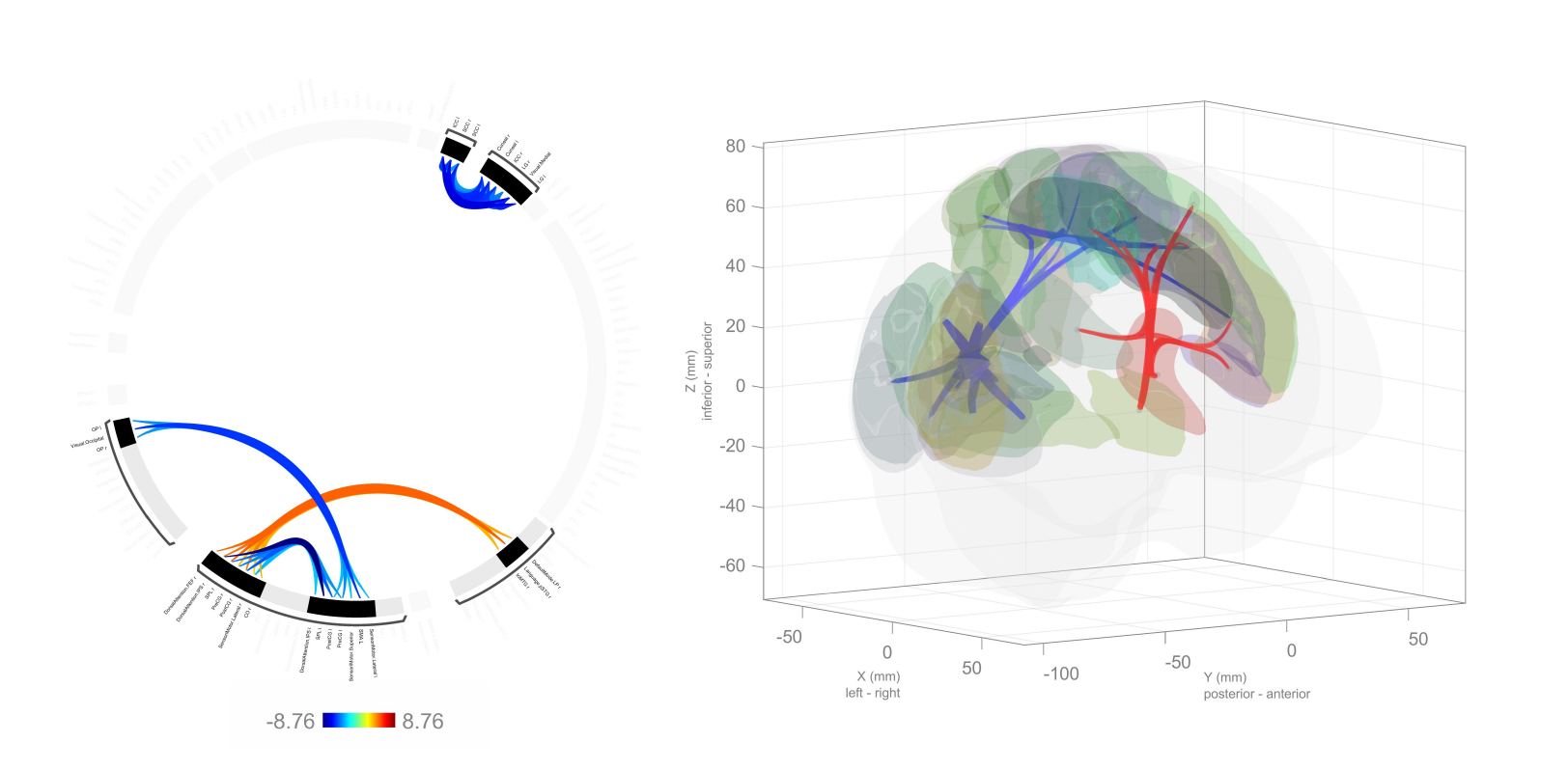}}
  \caption{Functional connectivity analysis results on quality assessment and content classification tasks. Shows the comparison of functional connectivity patterns in the task state and the resting state. Detailed brain network connectivity is shown on the left, the color bar on the left network diagram represents the t-value of the connection. Positive connectivity (positive correlation) is represented by the red line and negative connectivity (negative correlation) is represented by the blue line in the figure on the right.} 
  \label{conn} 
\end{figure*}

\subsection{fMRI Functional connectivity}
Based on the increased activity in the regions of the visual pathway in the QA task, it is reasonable to assume that the brain performs finer visual processing in the QA task, which may involve more complex processing compared to the CC task. We exploratively analyzed the functional connectivity of the QA task and the CC task and compared the similarities and differences in the functional connectivity of the two tasks. The results of the functional connectivity analysis are shown in Fig.\ref{conn}, which are both derived from task state and rest state contrasts.
Some connections are present in both QA tasks and CC tasks, such as the negative connection among the bilateral intra-calcarine cortex, supra-calcarine cortex, cuneus, and lingual gyrus. Another cluster of negative connections is in the motor region, including bilateral superior parietal lobule, postcentral gyrus, and precentral gyrus.

The CC task has two unique connections. The first is a negative connection between the bilateral occipital poles and the precentral and postcentral gyrus. The other is the positive connections between right-sided somatosensory, motor and attention-related brain regions (\textit{including the superior parietal lobule (SPL), precentral gyrus (PreCG), postcentral gyrus (PostCG), central opercular cortex (CO), inferior parietal lobule and lateral prefrontal region}) and right-sided language-, semantic-related brain regions (\textit{including superior temporal gyrus (pSTG) and middle temporal gyrus (toMTG)}).

It can be observed that QA tasks lead to richer functional connectivity. In addition to the above-mentioned, some important connections can be observed in QA tasks. The first is the negative connection between bilateral visual region (\textit{including the bilateral occipital cortex(iLOC), Temporal Occipital Fusiform Cortex (TOFusC)}) and somatosensory, motor, and attention-related brain regions (\textit{including the bilateral precentral gyrus (PreCG), postcentral gyrus (PostCG), Parietal Operculum Cortex(PO), superior parietal lobule (SPL), etc.}). Another one is a positive connection between the left and right sides of the four brain regions Middle Frontal Gyrus (MidFG), Frontal Pole (FP), Superior Frontal Gyrus (SFG), and lateral prefrontal cortex (LPFC), which belong to the higher cognitive regions.

We also analyzed functional connectivity in the high-quality and low-quality conditions and showed that there was no significant difference in functional connectivity in the brain between the two conditions.

\section{Discussion}
\subsection{Brain mechanisms of quality assessment tasks}
This study mainly investigates the impact of different image quality on the visual comprehension ability of the human brain and explores neural mechanisms related to visual quality assessment tasks. In the field of computer vision, people generally tend to classify image quality assessment as a low-level task\cite{wu2023q, zhang2011fsim}, while image semantic recognition tasks, such as image classification, are typically considered high-level tasks. However, we find that compared to content classification tasks, participants showed increased activation in bilateral visual pathways including the lingual gyrus, fusiform gyrus, cuneus, inferior temporal gyrus, and advanced cognitive regions such as inferior frontal gyrus, superior frontal gyrus, right insula when performing image quality assessment tasks.

The engagement of the visual accessory pathway, which is critical for processing complex visual information, executing goal-directed tasks, and controlling visual attention, suggests that image quality assessment tasks might demand more meticulous analysis and judgment of fine-grained visual attributes such as detail, clarity, color accuracy, contrast, among others, compared to content classification tasks. Moreover, these tasks necessitate the brain to process a greater amount of information related to the integrity and noise of visual signals, involving deeper processing within the visual cortex and related higher-order visual areas. Simultaneously, existing literature indicates that the superior frontal gyrus can coordinate working memory through low-frequency oscillations, playing an active role within it \cite{alagapan2019low}. The right insula, as part of the salience network, plays a significant role in modulating salience, a function that becomes particularly important in challenging tasks where attention may fatigue, leading to careless errors \cite{2019insular}. Furthermore, the Inferior Frontal Gyrus (IFG) is not only implicated in language production and comprehension \cite{tyler2005temporal}, but the left IFG is also thought to be involved in processes of inhibition, including the capacity to suppress learning from negative or detrimental information \cite{sharot2012selectively}.

In addition to the activations in visual pathways and advanced cognitive regions, more intricate functional connectivity patterns were observed during quality assessment (QA) tasks as compared to content classification (CC) tasks. Both tasks exhibited negative correlations between activity in visual-related cortices and brain regions associated with somatosensory, motor, and attentional regulation; however, QA tasks specifically engaged the inferior lateral occipital cortex (iLOC) and temporal occipital fusiform cortex (TOFusC) in this negative coupling, whereas CC tasks involved bilateral occipital poles.
The negative correlation between visual-related cortices and areas pertaining to somatosensory, motor, and attentional control suggests that, when executing these two types of tasks, the brain dynamically adjusts the activation states of different functional regions to accomplish visual processing tasks more efficiently. In the context of QA tasks, iLOC and TOFusC, acting as critical visual processing areas\cite{Grill2001lateral, kanwisher1997fusiform, mccandliss2003visual}, may need to suppress the allocation of non-directly related somatosensory, motor, or attention resources during fine-grained quality assessments to ensure focused attention on image quality details.
For CC tasks, where the bilateral occipital poles are implicated in this negative connection, content classification relies more heavily on recognizing overall shapes, colors, and basic features\cite{de2013anatomy}. The processing of such information is likely concentrated within primary and secondary visual cortices, particularly for rapid and straightforward analysis of basic visual information, thereby potentially reducing excessive engagement of the somatosensory, motor, and attention systems to make swift and accurate category judgments.
While both tasks demonstrate negative correlations between the visual system and other functional areas, the specific visual brain regions involved differ, reflecting distinct demands for optimized resource allocation in the brain according to the nature of the visual task at hand. It is reasonable to speculate that QA tasks emphasize higher-order and detailed visual perception, while CC tasks tend towards primary and rapid visual categorization — a perspective that aligns somewhat differently from the prevailing understanding in the computer vision field.

The QA task further reveals significant positive functional connectivity between homologous regions in the advanced cognitive areas of both hemispheres, including the middle frontal gyrus (MidFG), frontal pole (FP), superior frontal gyrus (SFG), and lateral prefrontal cortex (LPFC). This positive coupling suggests that during complex quality assessments, these high-level cognitive regions in the left and right hemispheres collaborate by sharing information and integrating analytical outcomes. Specifically, the middle frontal gyrus is involved in emotional and value-based judgments, the frontal pole possibly contributes to decision-making processes, the superior frontal gyrus relates to working memory and strategy selection, and the lateral prefrontal cortex is responsible for executive control and goal-directed behaviors\cite{boorman2013behavioral, chang2013neuronal, fitzgerald2009role, nicolle2012agent, noonan2011distinct}. The close coordination among these regions facilitates the formation of a comprehensive and integrated quality assessment conclusion and enables cross-hemisphere information exchange and comparison when necessary, ensuring the accuracy and consistency of the assessment process.

Based on the results of both univariate analyses and functional connectivity analyses in conjunction with QA and CC tasks, we can conclude that QA tasks are complex endeavors requiring not only heightened fine-grained visual perception but also intricate coordination among advanced cognitive regions associated with attention and working memory. The quality scores within extant datasets in the field of image quality assessment indeed stem from such a sophisticated cognitive process, which is distinct from the natural perceptual and cognitive mechanisms humans employ when encountering images. This distinction provides neuroscientific underpinnings for the subjective biases inherent in assessing image quality.

\subsection{Differences in brain response to high- and low-quality images}

The research findings indicate that there exist significant differences in the response of the brain's visual regions to images of varying quality when presented to the subjects: In the aforementioned results, when observers are confronted with high-quality images (characterized by high resolution, fine details, and well-organized structures), specific visual cortical areas such as the lingual gyrus, inferior occipital gyrus, and middle occipital gyrus exhibit stronger activation compared to their responses to low-quality images. These brain regions encompass the primary visual cortex BA17 and the secondary visual cortex BA18, which play pivotal roles in the chain of visual information processing. The BA17, also known as the primary visual cortex or V1 area, serves as the first stop for visual input, where it performs the initial encoding of basic visual features like lines, edges, and colors\cite{hubel1977ferrier, livingstone1988segregation, hubel1968receptive}. In contrast, the visual cortex containing BA18 engages more deeply in the processing of complex visual information, including shape recognition, object boundary integration, and understanding spatial relationships\cite{gattass1981v2, bakin2000visual}.

This suggests that these regions undertake more intricate and refined visual information processing tasks when handling high-resolution, detail-rich, and structurally coherent images, requiring the brain to engage in a more meticulous and profound analysis of these cues to achieve accurate identification and comprehension. The heightened activity levels observed in the secondary visual regions in response to high-quality images reflect the presence of richer overall shapes, fine textures, and three-dimensional structures among the visual features inherent in these images, which are more readily recognized by the human brain during perception.

When subjects are presented with low-quality images (such as those containing increased noise, blurriness, or distortions), certain higher-order visual regions, including parts of the fusiform gyrus, middle occipital gyrus, and superior occipital gyrus, display elevated levels of activation. These activations prominently involve Brodmann areas BA19 and BA37. BA19 is typically associated with complex shape recognition and depth perception\cite{larsson2006two, zeki1974functional}, whereas BA37, as a part of the fusiform gyrus, engages in advanced visual processing tasks such as scene comprehension\cite{patterson2007you, grill2006repetition, grill2004human}.

This suggests that when confronted with lower quality images or ones with more noise, blur, or distortion, the brain may allocate additional resources to compensate for missing information and to decode and identify visual content. That is, when the visual input quality is poor, the brain must overcome the extra challenges posed by noise, blurriness, and distortion to correctly interpret and make sense of the viewed material. By enhancing activity within these aforementioned regions, the brain might be endeavoring to extract useful information from the raw visual signals, attempting to reconstruct lost details, and conducting a deeper analysis of image content to facilitate object recognition and scene understanding.

Moreover, other regions, such as the middle temporal gyrus, inferior temporal gyrus, inferior frontal gyrus, middle frontal gyrus, superior parietal lobule, and inferior parietal lobule, also exhibit heightened levels of activation when observing low-quality images. Among these, the inferior temporal gyrus is particularly associated with object recognition and visual memory, playing a critical role in the storage and retrieval of long-term complex visual information\cite{Squire2007recognition, ranganath2012two, grill2014functional, haxby2001distributed}. The middle temporal gyrus, on the other hand, is involved in semantic memory and episodic memory processes, participating in language-related memory tasks\cite{Irish2013pivotal, xu2015tractography}.

This outcome reveals that when processing suboptimal visual input, the brain not only recruits relevant areas within the visual cortex but further mobilizes multiple brain regions including those governing attention regulation, cognitive control, and memory to compensate for missing information. It decodes and identifies visual content through intricate neural mechanisms. This phenomenon underscores the remarkable adaptability of the brain when faced with challenging visual conditions; it can flexibly adjust the operating modes of its neural networks to effectively process and utilize low-quality visual information.

In summary, the brain demonstrates dynamic adaptability when confronted with visual inputs of varying quality, adopting distinct strategies for processing information across different levels of visual fidelity. It adjusts its allocation of neural resources according to the quality of the input content, ensuring efficient processing and cognitive interpretation of the data. For high-quality images, it relies primarily on highly specialized visual regions for precise and detailed analysis; whereas for poor-quality images, it necessitates the mobilization of additional neural resources, including higher-order visual areas as well as associated cognitive and attentional regulatory networks, in order to effectively decode and recognize complex and ambiguous visual signals. This study, by comparing the differences in brain activity elicited by differing image qualities, provides empirical evidence for understanding how the brain adjusts its visual information processing strategies based on the varying quality of visual input.

\section{conclusion}
This study used fMRI to compare brain activity during Image Quality Assessment (QA) and Content Classification (CC) tasks. QA tasks were found to involve more complex functional integration, especially highlighting the role of the right iLOC and TOFusC in coordinating sensory, motor, and attentional resources for detailed quality analysis. Enhanced positive connectivity among high-level cognitive regions during QA tasks revealed interhemispheric cooperation for accurate and consistent quality judgments.

Furthermore, this research also comparatively analyzed the brain's responses to images of varying qualities. The findings demonstrate the brain's dynamic adaptability when encountering visual inputs of differing quality, adopting distinct strategies for processing visual information based on its fidelity. For high-quality images, the brain primarily relies on highly specialized visual areas for precise and detailed analysis. In contrast, for low-quality images, it mobilizes a broader array of neural resources, including higher-order visual regions and interconnected cognitive and attentional regulatory networks, to effectively decode and recognize complex and ambiguous visual signals.

In summary, this research, through fMRI data, elucidates how the brain adjusts regional activation based on image quality task demands. These findings contribute neurological evidence to understanding differing resource allocation across visual tasks and fill a knowledge gap in the influence of image quality on perception. This not only deepens our insight into human visual processing but also provides a scientific basis for refining IQA algorithms and exploring subjective assessment standards in computer vision.

\section{Limitations}
This study is the first to use fMRI to explore the brain perception mechanism of visual quality evaluation tasks, which is fundamental research in this field. This means that our experimental design has a coarse granularity and cannot cover all factors that affect visual quality. Due to the lack of a basic theoretical foundation in this field, our analysis at this stage mainly focuses on univariate analysis and functional connectivity analysis. Therefore, compared to visual fMRI datasets such as BOLD5000, we use fewer stimuli in our experiments, and the analysis of data-driven deep learning algorithms is limited. In the next stage, we will use multimodal stimulus materials, more types of loss, and finer grained visual quality to further explore the brain mechanism of visual quality perception. At the same time, we will also attempt to build a larger scale fMRI dataset for visual quality evaluation to better adapt to deep learning algorithms.
\clearpage
\bibliographystyle{IEEEtran}
\bibliography{IEEEabrv, reference.bib}
\end{CJK}
\end{document}